\documentstyle[prl,aps]{revtex}
\input epsf
\begin{document}
\draft
\title{ Search for solar axions using $^7$Li}
\author{M. Kr\v{c}mar$^{1}$\cite{MKrcmar}, 
        Z. Kre\v{c}ak$^{1}$, 
        A. Ljubi\v{c}i\'{c}$^{1}$,  
        M. Stip\v{c}evi\'{c}$^{1}$, and
        D. A. Bradley$^{2}$}
\address{$^{1}$Rudjer Bo\v{s}kovi\'{c} Institute, POB 180, 10002 Zagreb,
                 Croatia}
\address{$^{2}$School of Physics, University of Exeter, Stocker Road,
                 Exeter EX4 4Q1, UK}
\wideabs{
\maketitle
\begin{abstract}

We describe a novel approach to the search for solar, 
near--monochromatic hadronic axions,
the latter being suggested to be created in the solar core 
during $M$1 transitions 
between the first excited level of $^7$Li, at 478 keV, and the ground state. 
As a result of Doppler broadening, in principle these axions can be
detected via resonant absorption by the same nuclide on the Earth. Excited
nuclei of $^7$Li are produced in the solar interior by $^7$Be electron capture 
and thus the axions are accompanied by emission of $^7$Be solar 
neutrinos of energy 384 keV.
An experiment was made which has yielded an upper limit on 
hadronic axion mass of
32~keV at the 95\% confidence level.
\end{abstract}
}
\narrowtext 

Axions, neutral, spin--zero pseudoscalar particles that go beyond the 
Standard Model,
arise from spontaneous breaking of the Peccei--Quinn (PQ) chiral 
symmetry \cite{Pec77}, the latter being introduced to resolve the 
strong $CP$ problem. 
A non--zero axion mass ($m_a$) can be interpreted as a mixing of the axion 
field  with pions, and is related to the PQ symmetry
breaking scale ($f_a$) by 
$m_a f_a\!\approx\!m_{\pi} f_{\pi}$, where
$m_{\pi}\!=\!135$ MeV is the pion mass and $f_{\pi}\!\approx\!93$ MeV its
decay constant.
Generically, all effective coupling constants of axions with ordinary matter
and radiation are linear in $m_a$ or, equivalently, are inversely 
proportional to $f_a$.
The original suggestion that there existed axions, with $f_a$ equal 
to the scale of electroweak 
symmetry breaking and hence with $m_a$ of a few hundred keV, was quickly
ruled out by experiment. New axion models have subsequently 
been proposed which de--couple 
the PQ scale from the electroweak scale, and introduce $f_a$ at a value much
greater than 250~GeV. As such, the axion mass and all couplings become
extremely small and therefore axion models of this type 
are generically referred to
as invisible axion models. Two classes of invisible axion models have been
developed: KSVZ (Kim, Shifman, Vainshtein, and Zakharov) models 
\cite{Kim79} and DFSZ (Dine, Fischler, Srednicki, and Zhitnitski\u{i}) or grand 
unified theory (GUT) models \cite{Din81}. The main difference
between KSVZ and DFSZ axions is that the former have no tree--level
couplings to ordinary quarks and leptons because new heavy quarks have been
introduced that carry the PQ charge while usual quarks and leptons do not.
As a result, the interaction of KSVZ--type axions with electrons is 
strongly suppressed. In spite of this, their coupling to 
nucleons is not zero due to the generic
axion--pion mixing which exists even if the tree--level 
coupling to ordinary quarks vanishes. Since the KSVZ axions 
do not couple directly to leptons, they
are referred to as hadronic axions. The coupling of invisible axions
to photons is described by 
$g_{a\gamma\gamma} \propto m_a \{E/N\!-\!2(4\!+\!z\!+\!w)/
\,[3(1\!+\!z\!+\!w)]\}$, where the value of the 
parameter, $E/N$, is model dependent for hadronic axions, being a function
of exotic fermion charges, while the other parameter within the 
brackets is a function of quark mass
ratios $z\! \equiv\! m_u/m_d\! \approx\! 0.55$ and 
$w\! \equiv\! m_u/m_s\! \approx\! 0.029$
\cite{Leu96}. Note that in a recent review of the Particle Data Group 
$z$ was listed to have a value within the
conservative range 0.2 to 0.8 \cite{Pdg00}. It should be mentioned that
contrary to the KSVZ models, where $E/N$ is a model dependent number, in 
GUT models $E/N\!=\!8/3$.
It was shown by Kaplan \cite{Kap85} that it is possible 
to construct hadronic
axion models with $E/N\!=\!2$, in which the axion to photons coupling
is significantly reduced and may actually vanish because of a cancellation of
these two unrelated numbers. As such, the globular--cluster
constraints on axion mass \cite{Raf96},
which rule out DFSZ-type axions
with $m_a\!\agt\!10^{-2}~{\rm eV}$, as well as 
the most sensitive detection techniques for searching for
invisible axions \cite{Hag98,Oga96,Mor98,Zio99,Dil00}, based on the axion to
photons interaction, have no relevance to hadronic axions which only couple
to nucleons. So far the most restrictive boundaries on the 
mass of the hadronic axion arise from arguments concerning the
supernova (SN) 1987A cooling \cite{Raf96} and axion burst 
\cite{Eng90}. This narrow range of allowed axion masses, $10~{\rm eV}\!\alt\!
m_a\!\alt\!20~{\rm eV}$, is referred to as the hadronic axion window. 
Implications of nuclear axion emission in globular--cluster stars on limits
for axion mass, involving a metallicity dependent modification of the core
mass at helium ignition, have also been considered \cite{Hax91}. It should
be noted that supernova arguments depend upon axion emission in a hot and 
dense nuclear medium, one problem being to provide a reliable 
estimate of the axion emission rate from nucleon--nucleon bremsstrahlung, 
and on particle physics parameters where there exist large
uncertainties and ambiguities. As a result, the SN 1987A limits can be 
considerably relaxed \cite{Krc98f}.
In addition, the SN 1987A arguments suffer
from statistical weakness, with only 19 neutrinos being 
observed at the Kamiokande II and at the Irvine--Michigan--Brookhaven
water \v{C}erenkov detectors. Although supernovae of similar
size and distance to that of SN 1987A are very rare events, with only 
four historical records
over the past millennium, verification of hadronic axion window 
observations of another supernovae will be required if the uncertainty is to 
be reduced. In respect of the early universe, axions in the hadronic axion 
window can reach
thermal equilibrium before the QCD phase transition and hence, like neutrinos,
they are also candidates for hot dark matter \cite{Moro98}.

Because there exists a possibility that hadronic axions can be emitted during
magnetic nuclear transition, Moriyama \cite{Mor95} proposed the 
existence of near-monochromatic
axions of energy 14.4 keV, produced by nuclear emission from the first,
thermally excited level in $^{57}$Fe in the hot core of the Sun. The proposed
axions provide an ability to
open a new line of hadronic axion investigation independent
of the uncertainties in the axion--nucleon scattering cross section and 
modeling of supernovae. Contrary to the situation for supernovae, 
the Sun is the best known star, being well described by the 
Standard Solar Model (SSM) \cite{Bah00}, and
the solar axions are continuously available for experiments. 
The first experiment along this new line of solar axion
investigation set an upper limit on hadronic axion mass of 745 eV at the
95\% confidence level \cite{Krc98}. 

Referring to the thermonuclear fusion reactions that produce solar energy,
we propose a search for near--monochromatic hadronic axions of  
478 keV which might be emitted instead of $\gamma$ rays in the de--excitation
of $^7$Li. Towards this end, we refer to the particular reaction 
$^{3}\!He(\alpha,\gamma)^{7}\!Be$ and to the associated
reaction chains $^{7}\!Be(e^{-},\nu_e)^{7}\!Li(p,\alpha)\alpha$
and $^{7}\!Be(p,\gamma)^{8}\!B\!\rightarrow\!2 \alpha\!+\!e^{+}\!+\!\nu_e$ 
which generate
14\% and 0.1\% of the $\alpha$--particles, respectively, as well as 10.7\%
of the present--epoch luminosity of the Sun. The considerable amount of
excited $^7$Li nuclei which exist are produced by $^7$Be electron capture  
($^7\!Be\!+e^{-}\!\rightarrow\!^7\!Li^{\ast}\!+\!\nu_e$),
while a substantial flux of axions 
\begin{center}
     \begin{figure}[tbp]
      \epsfxsize = \hsize \epsfbox{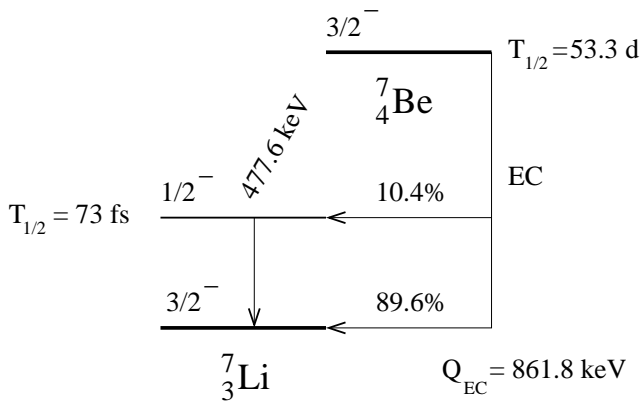}
      \caption{
        Decay scheme of $^7$Be. \label{fig1}}
   \end{figure}
\end{center}
   \noindent
are expected to accompany the emission 
of $^7$Be neutrinos of energy 384 keV. 
Figure \ref{fig1} shows the scheme of electron capture 
decay of $^7$Be \cite{Led78}. 
The branching ratio of the electron capture decay to the first excited state
of $^7$Li is $\kappa\!=\!0.104$. Note that the decay of the 478--keV excited 
level proceeds predominantly via $\gamma$--ray emission since the 
corresponding internal conversion
coefficient is $7.3\!\times\!10^{-7}$. Therefore the $\gamma$--decay 
width of the 
first excited to the ground state of $^7$Li ($\Gamma_{\gamma}\!=\!6.3\times\! 
10^{-6}$ keV) is a 
good approximation of the total decay width ($\Gamma$).
The high temperatures in the centre of the Sun ($\sim\!1.3$ keV) symmetrically
broaden the axion line to a full--width at half--maximum 
(FWHM) of about 0.5 
keV owing to the motion of the axion emitter. Since the effects
of nuclear recoil ($\approx\!1.8 \times 10^{-2}$ keV) and of redshift
due to the gravitation of the Sun ($\sim\!5\!\times\!10^{-3}$ keV) are much
smaller than Doppler shifts, the axion line is approximately centred at the
transition energy $E_{\gamma}\!=\!478$ keV. Because of thermal broadening
the near--monochromatic axions could be 
resonantly absorbed by the same nucleus
in a laboratory on the Earth, and the detection of subsequent emission of 
$\gamma$ rays of 478 keV would be a sign of axion existence.

As a result of the Doppler effect, the flux of axions accompanying
emission of $^7$Be solar neutrinos, differential with respect to axion 
energy $E_a$, $d\Phi (E_a)/dE_a$, has the particular form \cite{Met59}
\begin{eqnarray}
      \frac{d\Phi(E_a)}{dE_a} = \int_{0}^{R_{\odot}} && 
                                d\Phi_{\nu}^{Be}(r)\,
                                \kappa\,\frac{\Gamma_a}{\Gamma_{\gamma}}\,
                                \frac{1}{\sqrt{2\pi}\,\sigma(T)} 
      \nonumber \\
      && \times 
      \exp\left[-\frac{(E_a-E_{\gamma})^2}{2\,\sigma(T)^2}\right],
      \label{eq1}
\end{eqnarray} 
where $R_{\odot}$ denotes the solar radius, and $d\Phi_{\nu}^{Be}(r)$ is the
fraction of the $^7$Be neutrino flux at the Earth, produced in a 
given spherical shell in the solar interior at radius $r$.
The quantity $\sigma(T)\!=\!E_{\gamma}\,(k\,T/\,m)^{1/2}$ represents
Doppler broadening of the 478--keV axion line of $^7$Li at the temperature 
of the Sun ($T$) at the radius $r$, with $k$ and $m$ denoting the Boltzmann
constant and the mass of the $^7$Li nucleus, respectively. The branching ratio
($\Gamma_a/\Gamma_{\gamma}$) of the $M$1 axionic transition 
relative to the gamma
transition is calculated in the long--wavelength 
limit to be \cite{Hax91,Avi88} 
\begin{equation}
  \frac{\Gamma_{a}}{\Gamma_{\gamma}} = \left( \frac{k_a}{k_{\gamma}}
  \right)^{3} \frac{1}{2\pi\alpha}\; \frac{1}{1 + \delta^{2}} 
  \left[ \frac{g_{0}\beta + g_{3}}
  {\left( \mu_{0} - \frac{1}{2} \right)\beta + \mu_{3} - \eta} \right]^{2},
  \label{eq2} 
\end{equation}
where $k_{\gamma}$ and $k_{a}$ are the momenta of the photon and the axion,
respectively. $\alpha\!\approx\!1/137$ is the fine structure
constant, and $\delta\sim\!0$ is the $E$2/$M$1 mixing ratio.
The quantities $\mu_{0}\!\sim\!0.88$ and $\mu_{3}\!\sim\!4.71$ denote 
the isoscalar and isovector magnetic moments, respectively. The 
nuclear--structure--dependent terms $\eta$ and $\beta$ 
are estimated to have values 0.5 and 1, respectively, 
since the neutron shell of the $^7$Li is closed, and the present 
$M$1 transition is considered to be caused predominantly by the proton. The 
isoscalar ($g_0$) and isovector ($g_3$) axion--nucleon coupling constants 
as well as the axion mass are 
related to $f_a$ in the hadronic axion model \cite{Kap85,Sre85} by
expressions
 \begin{eqnarray}
      g_{0}&=&-\frac{m_N}{f_a}\,
                 \frac{1}{6}\,
                 \left[\,2S+\left(3F-D\right)\frac{1+z-2w}{1+z+w}\,\right],
      \label{eq3} \\
      g_{3}&=&-\frac{m_N}{f_a}\,\frac{1}{2}\,
                \left(D+F\right)\frac{1-z}{1+z+w}\;,
      \label{eq4}
\end{eqnarray}
and
 \begin{eqnarray}
        m_a &=& \frac{f_{\pi}\,m_{\pi}}{f_a}\,
                \left[\frac{z}{(1+z+w)(1+z)}\right]^{1/2}
    \nonumber \\
            &=& 6\,{\rm eV}\,\frac{10^6\,{\rm GeV}}{f_a}\;,
    \label{eqm}
 \end{eqnarray}
where $m_N\!\approx\!939$ MeV is the nucleon mass. 
The two axial--coupling parameters $F\!=\!0.460$ and $D\!=\!0.806$ 
are determined from hyperon semi--leptonic decays \cite{Rat96}. 
The flavor--singlet axial--vector matrix element ($S$) is
extracted from polarized structure function data, in a scheme dependent
way. This is still a poorly constrained parameter because the separation
between gluon and quark singlet contributions is, as usual, ambiguous beyond
leading order. In our calculations we have used a recently
estimated value from experimental data of the scale independent quark 
spin content of the nucleon $S\approx\!0.4$ \cite{Mal99}.

The rate of excitation per $^7$Li nucleus, which is expected for  
solar--produced axions incident on a laboratory target of $^7$Li is
 \begin{equation}
      R_N = \int_{-\infty}^{+\infty}dE_a\,\frac{d\Phi(E_a)}{dE_a}\,
            \sigma_D(E_a)\;,
   \label{eq5}
\end{equation}
where the effective cross section for resonant absorption of axions is 
given by \cite{Mor95,Met59}
\begin{equation}
     \sigma_D(E_a) = \sigma_0\,\frac{\Gamma_a}{\Gamma_{\gamma}}\,
                     \frac{\sqrt{\pi}\,\Gamma}{\sqrt{2}\,\sigma(T_E)}\,
            \exp\left[-\frac{(E_a-E_{\gamma})^2}{2\,\sigma(T_E)^2}\right], 
   \label{eq6}
\end{equation}
with Doppler broadening effects at room temperature ($T_E\!\sim\! 
300~{\rm K}$) allowed for, and 
$\sigma(T_E)\!\sim\!1~{\rm eV}\!\gg\!\Gamma$. The
maximum resonant cross section of $\gamma$ rays 
is expressed as $\sigma_0\!=\!
2\pi\,g\,\lambdabar^2\,\Gamma_{\gamma}\,/\,\Gamma\!=\!5.4\!\times\! 
10^{-21}~{\rm cm}^2$,
where $\lambdabar\!=\!\hbar\,c/E_{\gamma}$, $E_{\gamma}$ is the $\gamma$--ray 
energy, and $\hbar\,c\!=\!1.973\!\times\!10^{-8}~{\rm keV\,cm}$. The 
statistical weight factor $g\!=\!(2I_f+1)/(2I_i+1)$ and contains the 
total angular momenta 
$I_f$ of the excited level and $I_i$ the ground state value. 

Using equations (\ref{eq1}) for the differential axion flux at the Earth 
and (\ref{eq6}) for the resonant cross section in Eq.\ (\ref{eq5}), the
energy integration can be performed explicitly, leading to
\begin{equation}
     R_N = \sqrt{\frac{\pi}{2}}\,\kappa\,\sigma_0\,\Gamma\,\left(\frac
        {\Gamma_a}{\Gamma_{\gamma}}\right)^2
        \int_{0}^{R_{\odot}} \frac{d\Phi_{\nu}^{Be}(r)}{\sqrt{\sigma(T)^2
        + \sigma(T_E)^2}}\;.
   \label{eq7}
\end{equation}
Integrating this expression over the BP2000 SSM \cite{Bah00} which predicts
a total $^7$Be neutrino flux of $4.8\!\times\!10^{9}$ cm$^{-2}$\,s$^{-1}$,
and translating the rate of axionic excitation per 
$^7$Li nucleus into the total 
excitation rate, $R$, per unit mass of $^7$Li per day, we find  
\begin{equation}
     R = 1.3 \times 10^{-17} \left(\frac{m_a}{\rm 1 eV}\right)^4\;
         {\rm g}^{-1} {\rm day}^{-1}.            
   \label{eq8}
\end{equation}

We have searched for a peak corresponding to the 478--keV gamma ray of $^7$Li
in a single spectrum measured by a HPGe detector, having a crystal size of
about 50 mm$\phi \times$40 mm, 
at ground level. By using the veto NaI detector (8HW10/(4)3L, Bicron)
and iron--lead shielding the background events are 
reduced by a factor $\sim$10. We have obtained an energy calibration which 
identifies the gamma ray peaks that arise from environmental
radioactivity present. Energy resolution (FWHM) at 
the photon energy of 478 keV was
estimated to have a value of 1.4 keV. The target of lithium (Aldrich
Chemical Company),
\begin{center}
      \begin{figure}[tbp]
      \epsfxsize = \hsize \epsfbox{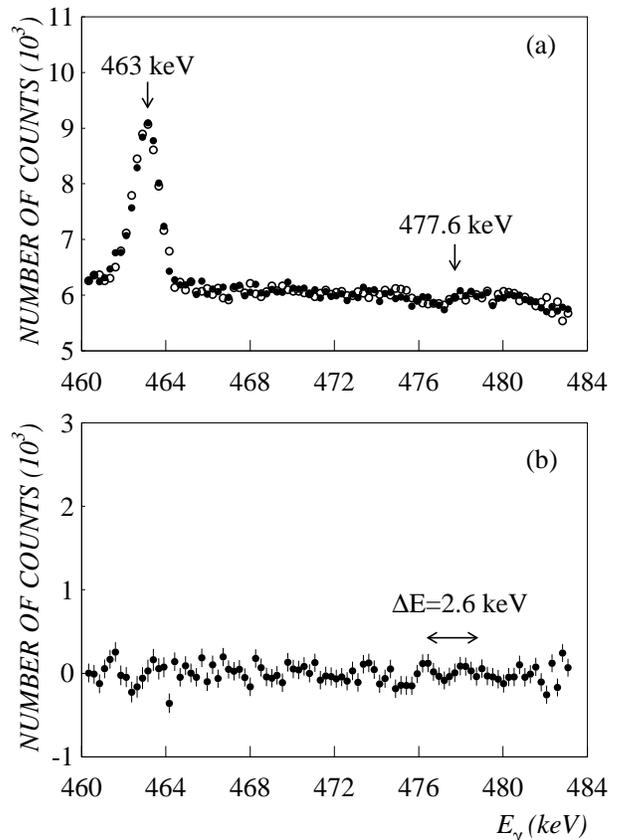}
      \caption{
        (a) The energy spectra accumulated for $9.6\times10^6$ s.
        Filled and open circles denote the run and background data,
        respectively. The peak at 463 keV arises from environmental
        radioactivity and corresponds to the $\gamma$ transition in
        $^{228}$Th. (b) Net number of counts (run $-$ background) in the
        region of the 478--keV gamma ray peak determined in an effort
        to detect axionic excitation of $^7$Li. \label{fig2}}
   \end{figure}
\end{center}
   \noindent 
an ingot with a diameter of 5.7 cm and a thickness of
4.5 cm, was placed 0.5 cm from the window of the detector. The mass of 
$^7$Li in the target is $M\!=\!56.72$ g. Data and background were both counted
for collection times $t\!=\!111.11$ days; gamma ray attenuation
due to the lithium ingot was taken into consideration. The
detection efficiency of the HPGe detector for 478--keV photons escaping from
the target was estimated to 
be $\varepsilon\!=\!8.3\!\times\!10^{-3}$; this was
evaluated by using standard bulk sources IAEA--0375 (soil) and IAEA--0373 
(grass) which contain the radionuclides $^{134}$Cs and $^{137}$Cs. 
The expected 
number of $\gamma$ rays detected in the HPGe detector is then given by 
$N_{\gamma}\!=\!R\,\varepsilon\,M\,t$, and depends on the fourth power of the
axion mass.

Figure \ref{fig2}(a) shows the energy spectra that have been measured with
the lithium target, providing sensitivity to the monochromatic solar axions,
and with an appropriate absorber to simulate background conditions.
The difference between these two spectra is shown in Fig.\ \ref{fig2}(b).
Centred at 477.6 keV, we have detected $N_{\gamma}\!=\!143\!\pm\!344$ 
photon events in the energy
interval of 2.6 keV. This yields an upper bound on the hadronic axion mass
$m_a\alt32~{\rm keV}$ at the 95\% confidence level. Note that we have
omitted a phase space factor $(k_a/k_{\gamma})^3$ in Eq.\ (\ref{eq2}); this 
affects the measured axion mass in our experiment by less than 
$4\!\times\!10^{-3}$. In the frame of SSM (maximal solar abundance of $^7$Li
by mass fraction $1.67\times 10^{-15}$ \cite{Tur93}) and by using the 
expression for the mean free path of the axions from Ref.\ \cite{Krc98}, we
found solar absorption of axions of
mass $\alt\!$ 32 keV to be insignificant ($<\!10^{-10}$), 
with implication of free escape from the solar interior \cite{nucl01}.

The obtained limit is about a factor 40 weaker than
the only other laboratory--evaluated limit for near--monochromatic
solar axions that has been  
made, the latter measurement concerning nuclear transition in 
$^{57}$Fe \cite{Krc98}. 
The present result has demonstrated that the newly--proposed source of
solar axions, being that associated with $^7$Be solar neutrinos, has 
the potential to provide the
parameter space of hadronic axion masses free from the large uncertainties
arising from SN 1987A considerations as well as from the uncertainty related
to the $^{57}$Fe-method with respect to determination of the iron 
abundance in the solar core. 
Namely, while the flux of $^{57}$Fe axions is
determined, in the frame of SSM, by extrapolation of 
the iron abundance in the solar photosphere
to the iron abundance in the centre of the Sun, 
the flux of $^7$Li axions is connected
with the $^7$Be solar neutrino flux \cite{SNO01}. This is to be
measured in a new experiment, 
Borexino \cite{Ali00}, via $\nu\!-\!e$ scattering
in 300 tonnes of liquid scintillator, with a potential detection 
threshold of as low as 250 keV. 
In addition, because of the possibility for improvements in detection 
of 478--keV gamma rays from $^7$Li atoms and further
background suppression, solar axion investigations with both,
$^7$Li and $^{57}$Fe, for probing the hadronic axion window, 
point towards a variety of data that are independent of supernova models,
their statistical weakness and the uncertainties associated with them.

The authors wish to thank the Ministry of Science and Technology of
Croatia for financial support. One of the authors (D.A.B.) is grateful
for financial support from the University of Exeter.


\begin{references}
      \bibitem[*]{MKrcmar}Corresponding author.\\Email address:
       mkrcmar@rudjer.irb.hr
      \bibitem{Pec77}
          R. D. Peccei and H. Quinn, Phys. Rev. Lett. {\bf
          38}, 1440 (1977); Phys. Rev. D {\bf 16}, 1791 (1977).
      \bibitem{Kim79}
          J. E. Kim, Phys. Rev. Lett. {\bf 43}, 103 (1979); M. A. Shifman,
          A. I. Vainshtein, and V. I. Zakharov, Nucl. Phys. {\bf B166},
          493 (1980).
      \bibitem{Din81}
          M. Dine, W. Fischler, and M. Srednicki, Phys. Lett. B
          {\bf 104}, 199 (1981); A. R. Zhitnitski\u{i}, Yad. Fiz.
          {\bf 31}, 497 (1980) [Sov. J. Nucl. Phys. {\bf 31}, 260
          (1980)].
      \bibitem{Leu96}
          H. Leutwyler, Phys. Lett. B {\bf 378}, 313 (1996).
      \bibitem{Pdg00}
          D. E. Groom {\it et al}., Eur. Phys. J. C {\bf 15}, 1 (2000). 
      \bibitem{Kap85}
          D. B. Kaplan, Nucl. Phys. {\bf B260}, 215 (1985).
      \bibitem{Raf96}
          See, e.g., G. G. Raffelt, {\it Stars as Laboratories for 
          Fundamental Physics}
          (The University of Chicago Press, Chicago, 1996).
      \bibitem{Hag98}
          C. Hagmann {\it et al}., Phys. Rev. Lett. {\bf 80}, 2043 (1998).    
      \bibitem{Oga96}
          I. Ogawa, S. Matsuki, and K. Yamamoto, Phys. Rev. D {\bf 53},
          R1740 (1996).
      \bibitem{Mor98}
          S. Moriyama {\it et al}., Phys. Lett. B {\bf 434}, 147 (1998).
      \bibitem{Zio99}
          K. Zioutas {\it et al}., Nucl. Instr. Meth. {\bf A425}, 480 (1999).
      \bibitem{Dil00}
          L. Di Lella, A. Pilaftsis, G. Raffelt, and K. Zioutas, 
          Phys. Rev. D {\bf 62}, 125011 (2000).
      \bibitem{Eng90}
          J. Engel, D. Seckel, and A. C. Hayes, Phys. Rev. Lett. {\bf 65},
          960 (1990).
      \bibitem{Hax91}
          W. C. Haxton and K. Y. Lee, Phys. Rev. Lett. {\bf 66}, 2557 (1991).
      \bibitem{Krc98f}
          The upper limit on hadronic axion mass has been recalculated to be
          $m_a \alt 40~{\rm eV}$ by taking only the current particle physics 
          parameters into account, see Ref.\ \cite{Krc98}.
      \bibitem{Moro98}
          T. Moroi and H. Murayama, Phys. Lett. B {\bf 440}, 69 (1998).
      \bibitem{Mor95}
          S. Moriyama, Phys. Rev. Lett. {\bf 75}, 3222 (1995).
      \bibitem{Bah00}
          See, e.g., J. N. Bahcall, M. H. Pinsonneault, and S. Basu,  
          arXiv.org e--Print archive, astro--ph/0010346 (2000). 
      \bibitem{Krc98}
          M. Kr\v{c}mar {\it et al}., Phys. Lett. B {\bf 442}, 38 (1998).
      \bibitem{Led78}
          {\it Table of Isotopes}, edited by C. M. Lederer and V. S. Shirley
          (Wiley--Interscience, New York, 1978), 7th ed.
      \bibitem{Met59}
          F. R. Metzger, Prog. Nucl. Phys. {\bf 7}, 54 (1959).     
      \bibitem{Avi88}
          F. T. Avignone III {\it et al}., Phys. Rev. D {\bf 37}, 618 (1988).
      \bibitem{Sre85}
          M. Srednicki, Nucl. Phys. {\bf B260}, 689 (1985).
      \bibitem{Rat96}
          P. G. Ratcliffe, Phys. Lett. B {\bf 365}, 383 (1996).
      \bibitem{Mal99} 
          G. K. Mallot, eConf C990809, 521 (2000).
      \bibitem{Tur93}
          S. Turck--Chi\`{e}ze {\it et al}., Phys. Rep. {\bf 230}, 57 (1993).
      \bibitem{nucl01}
          Referring to Ref.\ \cite{Led78}, we do not find any other nuclides
          (in the Sun or on the Earth), with nuclear states of similar energy 
          to that of 477.6 keV, which are able to resonantly absorb the solar
          axions.
      \bibitem{SNO01}
          The new experimental result of the SNO Collaboration
          confirms that the $^{8}$B solar neutrino flux is in excellent
          agreement with predictions of SSM, Q. R. Ahmad {\it et al}.,
          The SNO Collaboration,  
          arXiv.org e--Print archive, nucl--ex/0106015 (2001). 
      \bibitem{Ali00}
          G. Alimonti {\it et al}., Borexino Collaboration,
          arXiv.org e--Print archive, hep--ex/0012030 (2000).

\end{references}
\end{document}